\documentclass[a4paper,10pt]{article}

\pdfoutput=1 

\usepackage{jheppub} 

\usepackage[T1]{fontenc} 

\usepackage{setspace}

\usepackage{physics}
\usepackage{dsfont}

\usepackage[dvipsnames]{xcolor} 

\newcommand{\mB}{\mathcal{B}}
\newcommand{\mH}{\mathcal{H}}
\newcommand{\si}{\mathcal{I}}

\newcommand{\ronak}{}
\newcommand{\miguel}{}

\title{Scalar Asymptotic Charges and Dual Large Gauge Transformations}

\author[a]{Miguel Campiglia,}
\author[b]{Laurent Freidel,}
\author[b]{Florian Hopfmueller}
\author[b,c,d]{and Ronak M Soni}

\vspace{5mm}

\affiliation[a] {\it   Instituto de F\'isica, Facultad de Ciencias, \\ Igu\'a 4225, 
              esq. Mataojo, 11400 Montevideo, Uruguay.}
\affiliation[b] {\it Perimeter Institute for Theoretical Physics, \\
	31 Caroline Street North, Waterloo, ON, N2L 2Y5, Canada}
\affiliation[c] {\it Department of Theoretical Physics,
 Tata Institute of Fundamental Research,\\  Colaba, Mumbai, 400005, India}
\affiliation[d]{\it Stanford Institute for Theoretical Physics, 382 via Pueblo, Stanford, CA 94305, USA}

\vspace{1cm}

\emailAdd{campi@fisica.edu.uy}
\emailAdd{lfreidel@perimeterinstitute.ca}
\emailAdd{fhopfmueller@perimeterinstitute.ca}
\emailAdd{ronakms@stanford.edu}

\abstract{
In recent years soft factorization theorems in scattering amplitudes have been reinterpreted as conservation laws of asymptotic charges. In gauge, gravity, and higher spin theories the asymptotic charges can be understood as canonical generators of large gauge symmetries. Such a symmetry interpretation has been so far missing  for scalar soft theorems. We remedy this situation by treating the massless scalar field in terms of a dual two-form gauge field. We show that the asymptotic charges associated to the scalar soft theorem can be understood as  generators of large gauge transformations of the dual two-form field. 

The dual picture introduces two new puzzles: 
the charges have very unexpected Poisson brackets with the fields, and 
  the monopole term does not always have a dual gauge transformation interpretation.
  We find analogs of these two properties in the Kramers-Wannier duality on a finite lattice, indicating that the free scalar theory has new edge modes at infinity that canonically commute with all the bulk degrees of freedom.

}

\preprint{\parbox{3cm}{TIFR/TH/18-22}}

\begin{document} 
\maketitle
\flushbottom

\section{Introduction}
It has recently been discovered, see \cite{Strominger:2017zoo} for a review for gauge theories and gravity, that soft theorems of scattering amplitudes --- universal behaviour in the limit of low energy of one or more external massless particles --- are Ward identities of large gauge transformations --- gauge transformations that do not die off at the boundary of spacetime and are therefore real transformations.
The investigation of large gauge transformations has proven illuminating in other ways as well, most notably in connections to the memory effect \cite{Strominger:2014pwa} and a better understanding of the necessity of the Faddeev-Kulish dressing \cite{Kapec:2017tkm,Carney:2018ygh}.
They have also been understood in p-form \cite{pform1,pform2} and higher spin \cite{Campoleoni:2018uib} theories.

The connection between gauge transformations and soft theorems only 
makes sense when the soft particle is a gauge boson, like a photon, gluon or graviton.
However, soft theorems also exist (at tree level) in the theory of a massless scalar, which raises the question of whether scalar soft theorems are related to asymptotic symmetries.\footnote{\ronak{It should be mentioned that memory effects, which are  related to asymptotic symmetries in gauge theories, also continue to exist in scalar theories, see for example \cite{Satishchandran:2019pyc}.}}
With this motivation, Campiglia, Coito and Mizera \cite{Campiglia:2017dpg} calculated the charge whose Ward identity is the scalar soft theorem, in a four-dimensional theory in which the massless scalar has a three-point coupling with either a massive scalar or a massive fermion.\footnote{See also \cite{Campiglia:2017xkp} for the treatment of other even dimensions and \cite{Hamada:2017atr} for the treatment of pions.} They found that the charge in either case is given at future null infinity by
\begin{equation}
  Q^{+} [\lambda (\theta,\phi)] = 4\pi \int d\theta d\phi\ \lambda (\theta,\phi) \lim_{u \to -\infty} \lim_{r \to \infty} \left( r \phi (u,r,\theta,\phi) \right),
  \label{eqn:charge-intro}
\end{equation}
where $\phi$ is the massless scalar and the order of limits is important.
This charge can be written also as an integral over all of the future null and time-like infinities.
The part that is an integral over 
null infinity only involves the massless scalar (since the massive scalar is 
suppressed there) and is commonly known as the \emph{soft} part of the charge.
The part that depends on the massive field is called the \emph{hard} part.

Despite being able to write down the expression for the charge, 
they were unable to interpret it as a symmetry of the theory; in particular, assuming naive Poisson brackets makes it seem that the 
charge generates a transformation which is incompatible with the equations of motion.
It should be noted here that naive Poisson brackets are not always correct on null infinity \cite{Campiglia:2014yka}, and often correct transformations are given by modified Poisson brackets \cite{He:2014cra}. It is understood now that this modification is related  to the fact that phase space needs to be extended by edge modes degree of freedom and that this extension is needed to describe gauge theories with boundaries \cite{Donnelly:2016auv}. Thus, the above failure does not necessarily mean that the charges do not actually generate symmetries. 

In this paper, we propose an answer to this outstanding puzzle by relating this charge to large gauge transformations of a Hodge-dual two-form gauge field, \footnote{Asymptotic symmetries in higher-form theories have been discussed in \cite{pform1,pform2}. See also \cite{stringmemory} for related study of 2-form memory effect on strings.} 
given by
\begin{equation}
  d \mB = * d \phi.
  \label{eqn:duality-intro}
\end{equation}
More precisely, we show that the charge, when expressed in terms of two-form quantities, is exactly the generator of large gauge transformations, except when $\lambda$ is constant on the sphere.
In the case of constant $\lambda$, this relation only holds true when the scalar and gauge field are compact. 

While one may correctly object that Hodge duality only works for free fields, we note that this is not the first time that dual large gauge transformations have been shown to play a role in generating soft theorems.
In QED, it was shown that in a theory with only electric charges, the two polarisations of the soft photon were not independent \cite{He:2014cra}.
On adding magnetic charges and the associated large gauge transformations, however, the soft photon had two independent polarisations, one related to the electric large gauge transformations and one related to the magnetic ones \cite{Strominger:2015bla}, see also \cite{Hamada:2017bgi,Hosseinzadeh:2018dkh,Freidel:2018fsk}.
This is surprising, since a theory with both electric and magnetic charges cannot really be written in terms of a gauge field at all --- both $dF \neq 0$ and $d * F \neq 0$, so neither $F = dA$ nor $*F = d \tilde{A}$ makes sense.
The relation between large gauge symmetries and soft photons continues to work because large gauge transformations are real symmetries and should exist no matter how the theory is written down.
So, we may take the point of view that 
both the electric and magnetic large gauge charges are physical charges, and the Hodge duality is merely a technical way to understand their properties. 
This point of view was explored further for QED in \cite{Freidel:2018fsk} where it was shown that the presence of  edge modes allows an extension of the boundary symmetry algebra including not only the charge associated with the bulk gauge symmetry but also contains the dual symmetry generators.\footnote{\ronak{We do not expect to find the non-Abelian algebra found in \cite{Freidel:2018fsk} in this case, since, as we will see, all the charges in the scalar case are dual charges --- the non-trivial commutators are between electric and magnetic charges in that case.}}

As evidence for our proposal of considering the large gauge symmetries of the asymptotic two-form field, we show that deriving the action of the charge on the scalar from its action on the two-form field makes the action compatible with the equations of motion, solving the issue raised in \cite{Campiglia:2017dpg}.

At this point, however, we are left with two new puzzles.
First, 
the $\lambda = \text{const.}$ case is special in that it does not arise from a two-form large gauge transformation.
Second, there seem to be degrees of freedom at the past of future null infinity that commute with the massless scalar field everywhere.
While we are able to trace the origin of these properties, we are not able to show that they are true in an independent way to confirm their existence, where by independent confirmation we mean proving these properties to be true without using the two-form field.
However, we do show that the lattice scalar theory dual to a two-form theory on a finite lattice has boundary variables with these two properties, indicating that the variable that appears in \eqref{eqn:charge-intro} is an unexpected extra degree of freedom of this sort.

The organization of this paper is as follows.
In section \ref{sec:review}, we review the main results from \cite{Campiglia:2017dpg}, in particular the soft theorem, the associated charge, and the puzzles raised by it.
Then, we review the scalar/two-form Hodge duality in four dimensions in \ref{sec:duality} and use it in section \ref{sec:main-point} to show that the scalar charge is the generator of large gauge transformations for the two-form field; this section contains the main result of this paper.
Finally, in section \ref{sec:kw}, we show the lattice analog of the unexpected properties of the scalar charge.
Factors in expressions involving forms and Hodge duality are written down in appendix \ref{sec:conventions}.

\section{Scalar Asymptotic Charges} \label{sec:review}
In this section, we review the scalar soft theorem, its translation to a conserved asymptotic  charge, and the puzzles raised by this translation. We follow \cite{Campiglia:2017dpg}, to which the reader is referred for more details.
We consider a model consisting of a massless scalar $\phi$ interacting with a massive scalar $\chi$, with a Lagrangian
\begin{equation}
  \mathcal{L} = - \frac{1}{2} (\partial\phi)^{2} - \frac{1}{2} (\partial \chi)^{2} - \frac{1}{2} m^{2} \chi^{2} + \frac{g}{2} \phi \chi^{2}.
  \label{eqn:L}
\end{equation}
The cubic vertex leads to a tree level soft theorem in the quantum theory:\footnote{Beyond tree level, $\phi$ is expected to get a mass due to loop corrections, such that there is no soft limit to speak of. We focus on the tree level theory.}
\begin{equation}
  \mathcal{A}_{n+1} (p_{1}\cdots p_{n}; \omega q) \xrightarrow{\omega \to 0} \frac{g}{2 \omega} \sum_{i \in \chi} \frac{1}{p_{i} \cdot q } \mathcal{A}_{n} (p_{1}\cdots p_{n}) + O(\omega^{0}).
  \label{eqn:soft-thm}
\end{equation}
where $p_i$ label the momenta of massive particles and $\omega q$ is the null wave vector  of a massless excitation. 

To understand the associated charges, we will need to fix asymptotic expansions of the fields at null and time-like future infinity. 
The equations for the past infinities are similar, and we stick to future infinity in this paper.

Future null infinity $\si^{+}$ is approached by taking $r \to \infty$ while keeping $u = t - r$ constant. It is the $r = \infty$ slice of Minkowski space in the retarded coordinates
\begin{equation}
  ds^{2} = - du^{2} - 2 du dr + r^{2} q_{AB} dx^{A} dx^{B} .
  \label{eqn:retarded-coords}
\end{equation}
In this limit, the massive field $\chi$ dies off. The massless field $\phi$ has the asymptotic expansion
\begin{equation}
  \phi (u,r,x^{A}) = \frac{\varphi (u,x^{A})}{r} + \dots
  \label{eqn:phi-asymptotic},
\end{equation}
which can be obtained by assuming that, asymptotically, $\phi$ is given by the standard free field mode expansion and then using a saddle point approximation. 

Future time-like infinity $i^{+}$ can be reached by taking the limit $\tau \to \infty$ in the coordinates
\begin{equation}
  \tau = \sqrt{t^{2} - r^{2}}, \quad \rho = \frac{r}{\sqrt{t^{2} - r^{2}}}.
  \label{eqn:hyperbolic-coords}
\end{equation}
The metric becomes
\begin{align}
	d s^2 = - d \tau^2 + \tau^2 \left(\frac{d \rho^2}{1+\rho^2} + \rho^2 q_{AB} d x^A d x^B\right).
\end{align}
Assuming again, that $\chi$ and $\phi$ are given by free mode expansions in this limit,  their asymptotic behavior near timelike infinity  are obtained through a saddle point evaluation
\begin{align}
  - \frac{g}{2} \chi^{2} &= \frac{j (\rho,x^{A})}{\tau^{3}} + \dots \nonumber\\
 \phi &= \frac{\phi_{i^{+}}}{\tau} + { \ldots }
 \label{eqn:i-plus-asymptotics}
\end{align}
The data on future null and future timelike infinity are not independent: 
The equations of motion are
\begin{equation}
  \Box \phi = - \frac{g}{2} \chi^{2}.
  \label{eqn:phi-eom}
\end{equation}
In the following we denote the asymptotic value of the scalar field on 
$\si$ by 
\begin{equation}
  \varphi_\pm(x^A) =  \varphi (u= \pm\infty, x^{A}).
\end{equation}
Requiring that the value of $\varphi_+$ matches with the value of $\phi_{i^+}$ on  $i^{+}$ as $\rho \to \infty$, gives the matching condition
\begin{align}
  \varphi_+ (x^{A}) &= \frac{1}{4\pi} \int_{i^{+}} d^{3} Y \frac{j (Y)}{Y \cdot x},
 \label{eqn:matching-condition}
\end{align}
where $Y^{\mu} = (\sqrt{1+\rho^{2}}, \rho y^{A})$ is a representation of points on $i^{+}$, defined so that $ Y^2=1$, $x^{\mu} = (1, x^{A})$ is null,  and $d^{3} Y$ is the measure on the hyperboloid $i^+$.

In \cite{Campiglia:2017dpg}, it was shown that the soft theorem \ref{eqn:soft-thm} can be understood as the Ward identities of a family of conserved charges $Q^+[\lambda(x^A)]$. In terms of the asymptotic data, the charges take the form
\begin{align}
  Q^{+} [\lambda (x^{A})] &= 4 \pi \int d^2 S\  \lambda(x^{A}) \varphi_-( x^{A}) \label{eqn:soft-charge-1}\\
  &= - 4 \pi \int_{-\infty}^{\infty} du \int d^2 S\ \ \lambda(x^{A}) \partial_{u} \varphi (u,x^{A}) + \int d^2 S\  \lambda(x^{A}) \int_{i^{+}} d^{3} Y\frac{j(y)}{Y(y)  \cdot x} \nonumber\\
  &\equiv Q^{+}_{soft} + Q^{+}_{hard}.
  \label{eqn:soft-charge-2}
\end{align}
 Here, $d^2 S$ is the normalised measure on the unit sphere.  In the second line, we have integrated by parts in $u$ and used the matching condition to bring the charge into the standard form of a soft charge plus a hard charge.
The fact that there is a charge whose conservation is the soft theorem raises two related questions:

Firstly, is the charge a Noether charge for a symmetry? Many asymptotic charges are Noether charges of local symmetries, but here that interpretation is not immediate since the scalar theory has no local symmetries.

Secondly, how can the action of the charge be understood?  Passing from a charge to its action requires fixing the Poisson brackets. Using the naive Poisson brackets\footnote{We get the second line from the first line by integrating the first one, and usng the integration constant to make the RHS anti-symmetric, as in \cite{He:2014cra}.}
\begin{align}
  \{ \varphi(u,x^{A}), \partial_{\tilde{u}} \varphi (\tilde{u}, \tilde{x}^{A}) \} &= \frac{1}{2} \delta(u-\tilde{u}) \delta^{(2)} (x^{A} - \tilde{x}^{A}), \nonumber\\
  \{ \varphi (u,x^{A}), \varphi (\tilde{u}, \tilde{x}^{A}) \} &= -\frac{1}{4} \text{sign} (u - \tilde{u}) \delta^{(2)} (x^{A} - \tilde{x}^{A}).
  \label{eqn:naive-pbs}
\end{align}
The action of the charge seems to violate the matching condition  \eqref{eqn:matching-condition}, because its LHS transforms while the RHS doesn't. 

In the next two sections, we show that the solution to the first puzzle is that these soft charges generate large gauge transformations of a dual two-form gauge field, \emph{and} that postulating this automatically solves the second puzzle. 

\section{Scalar/Two-Form Duality in Four Dimensions} \label{sec:duality}
In $D$ spacetime dimensions, a free scalar is dual to a free Abelian $(D-2)$-form gauge field.\footnote{Usually, a compact scalar is dual to a gauge field with a $U(1)$ gauge group; however, in this paper, since the `charge' $g$ of the massive field is not quantised, there is no reason to believe that the scalar field is compact and so we may take it to be non-compact. 
}
In four dimensions, the dual gauge field $\mB$ is a two-form field, related to the scalar field by\footnote{We have expanded the differential geometric notation in appendix \ref{sec:conventions}.}
\begin{equation}
  d \mB = * d \phi \quad \text{or} \quad 3 \partial_{[\mu} \mB_{\nu\rho]} = \varepsilon_{\mu\nu\rho\sigma} \partial^{\sigma} \phi.
  \label{eqn:two-form-defn}
\end{equation}
We define the field strength
\begin{equation}
\mH \equiv d \mB.
\end{equation}
With these identifications, the Lagrangian for a free scalar becomes
\begin{equation}
  - \frac{1}{2} d\phi \wedge *d\phi = - \frac{1}{2} \mH \wedge *\mH = - \frac{1}{12} \mH_{\mu\nu\rho} \mH^{\mu\nu\rho},
  \label{eqn:lagrangian-duality}
\end{equation}
which is the expected Lagrangian for a two-form field, though not canonically normalized.\footnote{\ronak{Questions about whether this two-form theory has a well-defined variational principle etc are analysed in \cite{Henneaux:2018mgn}.}}

The two-form Lagrangian has a gauge-invariance
\begin{equation}
  \mB \to \mB + \alpha, \quad \alpha = d \beta
  \label{eqn:2form-g-inv}
\end{equation}
and a related global symmetry\footnote{These are also called large gauge transformations, but they are of a different sort than the large gauge transformations discussed elsewhere in the paper.}
\begin{equation}
  \mB \to \mB + \alpha, \quad d\alpha = 0 \text{ and } \alpha \neq d\beta, \quad \text{i.e. } \alpha \in H^{2} (M),
  \label{eqn:2form-large-g-inv}
\end{equation}
where $M$ is the manifold on which the theory lives.
For our purposes, it will be more useful to work with $\alpha$ rather than $\beta$, for two reasons.
The first is that parametrising gauge transformations by $\beta$ requires one to deal with gauge-of-gauge redundancies, since $\beta$ and $\beta + d\gamma$ parametrise the same transformation of the gauge field $\mB$.
Secondly, since in the end we are interested in large gauge transformations, which are after all physical transformations, 
there will be no point distinguishing 
{\miguel them from global transformations.}

{\ronak Here, it is useful to stop and clarify some terminology.
  The transformations in \eqref{eqn:2form-large-g-inv} are global symmetries even in the theory without any boundaries, which can be seen by the fact that they modify the values of the Wilson lines.
  In a theory with boundaries, there are more global  symmetries having to do with the fact that the gauge parameter need not die off at the boundary, which are what are called large gauge transformations or asymptotic symmetries in our paper.
  Further restricting to Minkowski, this second class of large gauge transformations are what give rise to soft theorems.
  In the cases where the gauge parameter is a constant at infinity, they are very well-known as transformations which do not affect the gauge field itself but do affect any matter coupled to the gauge field\footnote{Note that this description does not apply to the two-form transformations in \eqref{eqn:2form-g-inv} with $\beta = d \gamma$, as these do not affect any matter coupled to the field.} --- examples are the gauge transformation with a spacetime-independent parameter in QED (whose conservation is the conservation of charge), and {\miguel asymptotic} Killing symmetries in gravity (whose conservations are the conservations of energy, momentum and angular momentum).
  The recent advances have been due to the recognition of the non-triviality of such large gauge transformations even when the gauge parameter is not constant at infinity, and the observation that the conservation laws associated to these symmetries are the soft theorems.
}

Strictly speaking, the duality works only for free fields: It exchanges the sourceless equations of motion with the Bianchi identity,
\begin{align}
  d*d\phi \propto \Box \phi = 0 &\Leftrightarrow d^{2} \mB = 0 \nonumber\\
  d^{2} \phi = 0 &\Leftrightarrow d*d \mB \propto \nabla^{\mu} \mB_{\mu\nu} = 0.
  \label{eqn:eom-bianchi}
\end{align}
This step fails if there are sources. However, as in electromagnetic duality, we may interpret a source for $\phi$ as a magnetic monopole for $\mB$.  This suggest a description of the massive $\chi$ particles in terms of worldlines as we now describe.\footnote{An alternative approach is to keep the field description for the massive particles and implement the duality only at $\si$.}
Calling the free action in \eqref{eqn:lagrangian-duality} $S_{B}$, we may write the full action as 
\begin{equation}
  S = S_{B} + S_{pp} + S_{int}.
  \label{eqn:action-terms}
\end{equation}
$S_{pp}$ is the free part of the massive scalar action, which may be written in the world-line picture as
\begin{equation}
  S_{pp} = - m \sum_{i} \int d \tau_{i},
  \label{eqn:s-pp}
\end{equation}
where the $i$'s { label the different particles and the integral is over the wordline of each particle.}

In terms of $\phi$, the interaction term $S_{int}$ is
\begin{equation}
  S_{int} = \frac{g}{2m} \sum_{i} \int d \tau_{i} \phi (x_{i} (\tau_{i})).
  \label{eqn:s-int-phi}
\end{equation}
To write it in terms of the two-form, we introduce another integral over the worldline
\begin{align}
  S_{int} &= \frac{g}{2m} \sum_{i} \int d \tau_{i} \int^{\tau_{i}} d \tau' \frac{d x^{\mu}}{d \tau'} \partial_{\mu} \phi (x_i(\tau')) \nonumber\\
  &= \frac{g}{12m} \sum_{i} \int d \tau_{i} \int^{\tau_{i}} d \tau' \frac{d x^{\mu}}{d \tau'} \varepsilon_{\mu\nu\rho\sigma} \mH^{\nu\rho\sigma} (x_i(\tau'))
  \label{eqn:s-int-B}
\end{align}
This expression demonstrates that the interaction can be written down in two-form language, and that it is non-local and gauge invariant.\footnote{\ronak{It should be noted that there are two reasons for the oddity of this interaction. First, two-form fields naturally couple to string-like charges, and trying to locally couple them to point-like charges necessarily leads one to the two-form field coupling to a total derivative, see for example \cite{DiGrezia:2003fe}. The non-locality comes from the fact that it is coupled to a source for the Hodge-dual field, and would remain in for example electric-magnetic duality where both sides of the duality naturally couple to point-like charges.}}

Of course, the world lines of the massive particles do not reach $\si$. That enables us to use the duality at $\si$ as if the theory were free, and consider the world line contributions only in the ``hard'' integrals over $i^+$. 

\section{Two-Form Asymptotic Symmetries} \label{sec:main-point}
\subsection{Fall-offs}\label{subsec:falloffs}
We now work out the charges corresponding to large gauge transformations for the two-form field, and show that it is the same as the scalar charge \eqref{eqn:soft-charge-1}.
By the falloffs of the scalar \eqref{eqn:phi-asymptotic} and the duality \eqref{eqn:two-form-defn}, the field strength for a two-form field has the null infinity fall-offs
\begin{align}
  \mH_{urA} &= \frac{H_{urA}^{(-1)}}{r} + \cdots \nonumber\\
  \mH_{uAB} &= r H_{uAB}^{(1)} + \cdots \nonumber\\
  \mH_{rAB} &= H_{rAB}^{(0)} + \cdots.
  \label{eqn:H-falloffs}
\end{align}
Working in Lorenz gauge
\begin{equation}
  \nabla^{\mu} \mB_{\mu\nu} = 0,
  \label{eqn:lorenz-gauge}
\end{equation}
the fall-offs \eqref{eqn:H-falloffs} can be obtained by demanding
\begin{align}
  \mB_{ur} &= \frac{B_{ur}^{(-1)} (x^{A})}{r} + \frac{B_{ur}^{(-2)} (u,x^{A})}{r^{2}} + \cdots \nonumber\\
  \mB_{uA} &= B_{uA}^{(0)} (u,x^{A}) + \cdots \nonumber\\
  \mB_{rA} &= B_{rA}^{(0)} (x^{A}) + \frac{B_{rA}^{(-1)} (u,x^{A})}{r} + \cdots \nonumber\\
  \mB_{AB} &= r B_{AB}^{(1)} (u,x^{A}) + \cdots
  \label{eqn:B-falloffs}
\end{align}
The pieces $B_{ur}^{(-1)}$ and $B_{rA}^{(0)}$ have to be $u$-independent to be compatible with the falloffs \eqref{eqn:H-falloffs} for $\mH$.
In fact, even setting the pieces $B_{ur}^{(-1)}$ and $B_{rA}^{(0)}$ to zero is consistent with a general $\mH$ with the right fall-offs. But as we will see, those boundary conditions would be too restrictive to accommodate the symmetry that gives rise to the scalar charges.

Equation \eqref{eqn:B-falloffs} leads to the identifications
\begin{align}
  H_{urA}^{(-1)} &= \partial_{u} B_{rA}^{(-1)} + \partial_{A} B_{ur}^{(-1)} \nonumber\\
  H_{uAB}^{(1)} &= \partial_{u} B_{AB}^{(1)} \nonumber\\
  H_{rAB}^{(0)} &= B_{AB}^{(1)} + 2 \partial_{[A} B_{B]r}^{(0)}.
  \label{eqn:H-B-identification}
\end{align}
Further, matching with scalar asymptotics gives the identification
\begin{eqnarray}
  \sqrt q \epsilon_{AB} \varphi = H_{rAB}^0 = B_{AB}^{(1)} + 2 \partial_{[A} B_{B]r}^{(0)} \qquad \Rightarrow \varphi = \frac{1}{\sin \theta} \left\{ B_{\theta\phi}^{(1)} + 2 \partial_{[\theta} B_{\phi]r}^{(0)} \right\},
  \label{eqn:phi-B-identification}
\end{eqnarray}
where we have gone to standard coordinates on the sphere for ease of notation.

Allowed residual gauge transformations are two-forms $\alpha$ that satisfy $d\alpha = 0$ and $\nabla^{\mu} \alpha_{\mu\nu} = 0$ and have the same or faster falloffs  than those of $\mB$.
We see, from the second and third lines of \eqref{eqn:H-B-identification}, that $d \alpha = 0$ implies
\begin{align}
  \alpha_{AB}^{(1)} (u,x^{A}) &= \alpha_{AB}^{(1)} (x^{A}) = -2 \partial_{[A} \alpha_{B]r}^{(0)} (x^{A}).
  \label{eqn:alpha-constraints}
\end{align}
From the second line, we see that demanding $B_{rA}^{(0)} = 0$ would have forced $\alpha_{AB}^{(1)}$ to vanish.\footnote{ In the Lorentz gauge 
$B_{ur}^{(-1)}$ is determined by the  divergence of 
$B_{rA}^{(0)}$  and required to stay in Lorenz gauge,  and same with the components of $\alpha$.}
Further, we also note for future reference that $\alpha_{AB}^{(1)}$ is necessarily exact on the celestial sphere.

\subsection{Charge}
Now, we may calculate the charge corresponding to these large gauge transformations. Let us first restrict to the free two-form theory, and use the symplectic form\footnote{Antisymmetrization of the $\delta$'s is understood.}
\begin{equation}
  \Omega = \int du d^{2} x\ \delta \mB \wedge *d \delta \mB = -\int du d^{2}x\ \frac{1}{\sin \theta} \delta B_{\theta\phi}^{(1)} \partial_{u} \delta B_{\theta\phi}^{(1)}.
  \label{eqn:two-form-symp-form}
\end{equation}
The soft part of the charge that generates the transformation $\mB \rightarrow \mB + \alpha$  is therefore given by  
\begin{align}
  Q_{soft} &= \int_{\si^+} \alpha \wedge *\mH = \int du d^{2}x\ [ \alpha_{\theta\phi} (*\mH)_{u} + \alpha_{u\theta} (*\mH)_{\phi} - \alpha_{u\phi} (*\mH)_{\theta}] \nonumber\\
 &= - \int du d^{2}x\  \alpha_{\theta\phi}^{(1)} \frac{\partial_{u} B_{\theta\phi}^{(1)}}{\sin \theta} = - \int d^{2} x  \frac{1}{\sin \theta} \alpha_{\theta\phi}^{(1)} [B_{\theta\phi}^{(1)}]_{-\infty}^{\infty} \nonumber\\
 &= - \int d^{2} x\ \alpha_{\theta\phi}^{(1)} [\varphi]^{\infty}_{-\infty}.
 \label{eqn:charge}
\end{align}
The terms proportional to $\alpha_{uA}$ disappear because they are subdominant, of order $1/r$.
Alternatively, the charge can be derived in form language:
\begin{align}
  Q_{soft} &= \int_{\si^{+}} \alpha \wedge *\mH = - \int_{\si^{+}} \alpha \wedge d \phi = - \int_{\si^{+}} d (\alpha \phi) \nonumber\\
  &= - \int d^{2} x\ \alpha_{\theta\phi}^{(1)} [\varphi]^{\infty}_{-\infty}.
  \label{eqn:charge2}
\end{align}
Recall that $\mH = d \mB$, and that $d \alpha = 0$.
{ \ronak Note here that this is a completely gauge-independent derivation of the main result: all that went into it is the matching of boundary conditions.}

Thus, we see that we have exactly reproduced the soft part of the scalar charge (which  in the free theory coincides with the total charge).

To include interactions, we have to include the fields at future infinity and take into account the transformations of the massive $\chi$ particles.  Despite the fact that the interaction of $\chi$ with the two-form is not local, the wordline picture gives us a neat way to do this.
Normally, to calculate the charge we would  perform the integral
\begin{equation}
  \int_{\Sigma_{3}} \alpha \wedge * \mH
  \label{eqn:charge-int}
\end{equation}
over an entire Cauchy surface, for example $\si^{+} \cup i^{+}$. Since at the world lines the two-form variables break down, we must however treat the world lines separately. Let $U_i$ be a small 3-ball around the point where the $i$th world line intersects $i^+$. 

Outside the regions $U_i$, the charge can be computed as
\begin{align}
Q_{\text{outside}} = \int_{\si} \alpha \wedge * \mH + \int_{i^+ - \cup_i U_i} \alpha \wedge *\mH. 
\end{align}
To compute the charge contribution from the excluded regions $U_i$, we can use that the charge density can be rewritten in terms of the scalar field:
\begin{equation}
\alpha \wedge * \mH = d ( \phi \, \alpha ).
\label{eqn:charge-is-tot-dvtv}
\end{equation}
Since the scalar variables continue to make sense at the world lines, we can integrate this over the regions $U_i$ to get the world line contributions to the charge. Using Stoke's theorem, the full charge is then
\begin{equation}
Q = \int_{\si^{+}} \alpha \wedge *\mH + \int_{i^{+} - \cup_{i} U_{i}} \alpha \wedge *\mH + \int_{\cup_{i} \partial U_{i}} \phi \, \alpha.
\label{eqn:full-charge}
\end{equation}
The last two terms can be rewritten as an integral over the future of $\si ^+$, by using \eqref{eqn:charge-is-tot-dvtv} again and keeping track of the orientations of $\partial U_i$:
\begin{align}
Q_{hard} = \int_{i^{+} - \cup_{i} U_{i}} \alpha \wedge *\mH + \int_{\cup_{i} \partial U_{i}} \phi \, \alpha = \int_{u = \infty} \phi \, \alpha  = \int d^{2}x   \alpha_{\theta\phi}  \varphi_+.
\label{eqn:hard-charge-on-ipp}
\end{align}
This expression is exactly the hard charge of the scalar theory! It continues to makes sense even if we consider $\chi$-configurations which are not localized to small regions, but spread out over the entirety of $i^+$.
 
Further, adding this to the soft charge \eqref{eqn:charge} we find the full charge associated with the asymptotic transformation $\alpha_{\theta\phi} (x^{A}) = 4\pi \sin \theta \lambda (x^{A})$ to be
\begin{equation}
Q^{+}[\lambda] = 4\pi \int_{S^{2}} \lambda (x^{A}) \varphi_- (x^{A}),
\label{eqn:full-charge-from-two-form}
\end{equation}
which is the same as the scalar charge \eqref{eqn:soft-charge-1}.

Before going ahead, we note that the worldline picture was 
unnecessary for the derivation of the hard charge.
We could have just calculated the Lienard-Wiechert potential for the massive field 
 to get equation \eqref{eqn:hard-charge-on-ipp} without any of the intermediate steps, similar to what was done in \cite{Kapec:2015ena}.
We used the wordline picture because it provided a nice physical interpretation,  because it makes clear that we do not know how the gauge transformations act on the massive field, and most importantly because it is required for finite time formulations of the charge where there is no clean separation between null and time-like regions.

Now that we have shown that the scalar charges are two-form large gauge charges, we turn to a couple of important subtleties of this identification.
First, in section \ref{ssec:pbs}, we investigate the symmetry action of these charges and its compatibility with the equations of motion.
Second, we discuss the missing charge; naively, because of the exactness condition, the case $\alpha_{\theta\phi} = \text{const.}$ is not an allowed large gauge transformation, and so it seems that \emph{all but one} of the scalar charges have been reproduced.
Finally, in the next section, we will show that the entire story here has an analog in Hamiltonian Kramers-Wannier duality.

\subsection{Understanding the Symmetry Action} \label{ssec:pbs}
Now that we have identified the scalar asymptotic charges as two-form large gauge charges, thus answering the puzzle of the existence of such a charge without a corresponding gauge symmetry raised in section \ref{sec:review}, we turn our attention to understanding the symmetry action of these charges on the scalar variables.

The main thing to note is that by the duality relation \eqref{eqn:two-form-defn}, any transformation that leaves the two-form field strength $\mH$ invariant also leaves the scalar invariant.
In particular, this means that the transformation generated by our asymptotic charges leave $\varphi (u,x^{A})$ invariant, so that
\begin{equation}
  \{ Q [\lambda], \varphi(u,x^{A}) \} = 0.
  \label{eqn:charge-real-action}
\end{equation}
This is because the charge generates a two-form transformation of the form \eqref{eqn:2form-g-inv}, so that the field strength is unaffected despite the transformation being a physical symmetry.

This in particular means that both sides of the matching condition \eqref{eqn:matching-condition} are left invariant by the asymptotic charges, making the symmetry action compatible with the equations of motion.
The second puzzle raised in section \ref{sec:review} has been automatically solved by the solution to the first one!
We do leave the question of whether it is a symmetry of the action unresolved, however, because we have not tried to understand exactly how the massive field transforms.

Note also that \eqref{eqn:charge-real-action} is starkly different from the action we would have predicted from the naive Poisson brackets \eqref{eqn:naive-pbs}, 
\begin{equation}
  \{ Q[\lambda], \varphi (u,x^{A}) \} = \pi \lambda (x^{A}).
  \label{eqn:charge-fake-action}
\end{equation}
Thus, we've found that --- as far as the free scalar theory is concerned --- these charges commute with everything and are thus in the centre of the algebra.
While this is not surprising from  the two-form point of view, it is very surprising from the point of view of naive expectations about the scalar field theory in Minkowski space: it seems that there are new `edge modes' on the boundary of the scalar theory that do not participate in dynamics.\footnote{We caution here that we do not understand the transformation of the massive field under the asymptotic charges and thus do not understand the correct version of these statements in the interacting theory; equally, \eqref{eqn:charge-fake-action} is surprising even in that case.}

\subsection{The Missing Charge} \label{ssec:0-mode}
The other peculiarity of the identification \eqref{eqn:charge} is that it provides a large gauge transformation interpretation for all the scalar asymptotic charges but one:  Recall from \ref{subsec:falloffs} that the asymptotic conditions forced the two-form gauge parameter $\alpha$ to be exact on the celestial sphere, while the gauge parameter for the scalar charges has to satisfy no such condition. Hence the scalar charge with $\alpha_{\theta\phi} = \sin \theta$ has no interpretation in the two-form theory. This is because $\sin \theta$, which is the area element of the unit two-sphere, cannot be written as a total derivative  of a smooth one-form field, as can be seen from the fact that its integral is non-zero.

For the case of a compact scalar/two-form field, this  can be resolved by allowing the one-form $\beta$ to wind around the  internal $U(1)$.
For example, taking
\begin{equation}
  \beta = - \phi \sin \theta d \theta \quad \Rightarrow d \beta = \sin \theta d\theta \wedge d \phi,
  \label{eqn:winding-beta}
\end{equation}
gives the desired value.
Of course, there needs to be a prefactor in this equation corresponding to the size of the $U(1)$ circle, or the coupling, and we have ignored it.

For the non-compact case, however, this cannot be done.\footnote{{ It is interesting to note that} 
the non-compact case may not even be well-defined \cite{Klebanov:1988ba,Konishi:1989wk},  { and so perhaps one should not} 
worry about it too much.} { We will see  in section \ref{sec:kw} an analogue of this situation on the lattice.} 

\section{Analogous Properties from Kramers-Wannier duality} \label{sec:kw}
We will now show that a free two-form field on a finite lattice is dual not  to the usual scalar theory, but a scalar theory with a centre, a set of variables that canonically commute with every variable in the theory; this sort of observation has also been made in \cite{Casini:2014aia,Radicevic:2016tlt,Moitra:2018lxn} in the context of entanglement entropy calculations.
Further, we show that all but one of these centre variables are the generators of large gauge transformations.
These are both exactly the peculiarities the soft charge \eqref{eqn:soft-charge-1}.
Thus, despite all the differences between a constant-time slice of a free theory on a finite lattice and null infinity of an interacting theory in Minkowski space, we propose that the variables $\varphi_-(x^a)=\varphi(u=-\infty,x^{A})$ that turn up in the charge \eqref{eqn:soft-charge-1} are exactly of this type.

Let us first review Kramers-Wannier duality \cite{Kramers:1941zz,Kramers:1941kn,Kogut:1979wt} between a two-form and a scalar.
We will work in the Hamiltonian formalism, by going to the analog of $B_{0i} = 0$ gauge and defining position-like and momentum-like variables on a three-dimensional constant-time slice.
The gauge fields of the two-form theory live on faces of the lattice, and they have an orientation, so that
\begin{equation}
  B_{i,\hat{\alpha},\hat{\beta}} = - B_{i,\hat{\beta},\hat{\alpha}} = -B_{i+\hat{\alpha},-\hat{\alpha},\hat{\beta}} = B_{i+\hat{\alpha}+\hat{\beta},-\hat{\alpha},-\hat{\beta}}.
  \label{eqn:B-two-form}
\end{equation}
where the face is labelled by a lattice point $i$ and the orientation of the two sides, here $\hat{\alpha},\hat{\beta}$, and the relative sign is fixed by the cross-product of the two orientations.
Similarly, each face also has a canonically conjugate variable, which we call $E_{i,\hat{\alpha},\hat{\beta}}$, with the same sign conventions as for $B$.
The Poisson brackets are
\begin{equation}
  \left\{ B_{i,\hat{\alpha},\hat{\beta}}, E_{i',\hat{\alpha'},\hat{\beta'}} \right\} = \delta_{(i,\hat{\alpha},\hat{\beta}),(i',\hat{\alpha'},\hat{\beta'})},
  \label{eqn:two-form-PBs}
\end{equation}
where the $\delta$ symbol is non-zero when the two faces are the same and has positive or negative sign depending on whether the orientations are the same or opposite.

\begin{figure}[h]
  \centering
  \includegraphics[width=50mm]{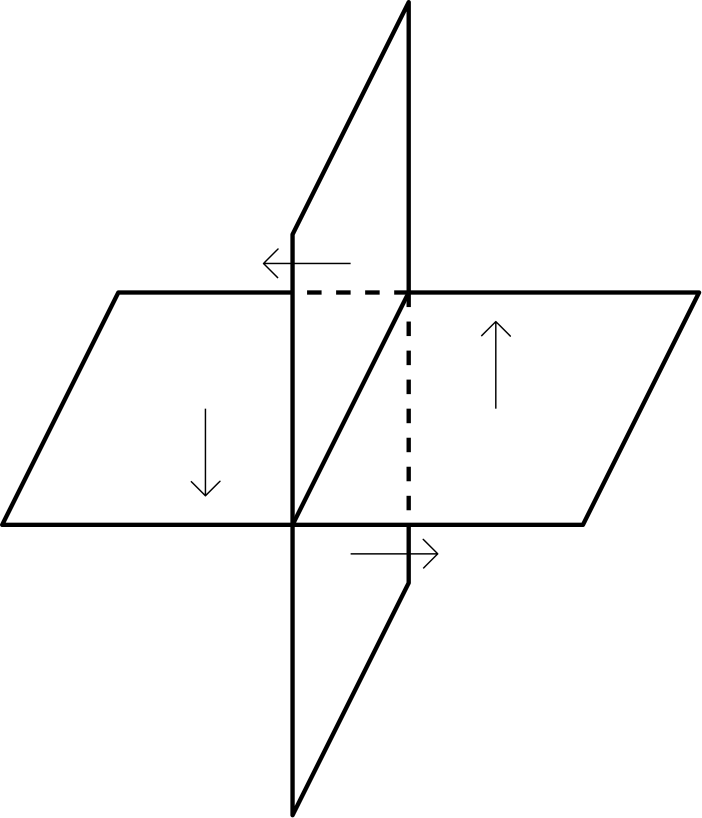}
  \caption{The local generator of the gauge transformations. The $E$s are added up on all four plaquettes adjoining a link with orientations given by the arrows.}
  \label{fig:g-tr}
\end{figure}

The gauge transformations are parametrised by `one-forms' that live on links,
\begin{equation}
  \alpha_{i,\hat{\alpha}} = - \alpha_{i+\hat{\alpha},-\hat{\alpha}}.
  \label{eqn:gauge-param}
\end{equation}
The gauge transformation is
\begin{equation}
  \delta B_{i,\hat{\alpha},\hat{\beta}} = \alpha_{i+\hat{\alpha},\hat{\beta}} - \alpha_{i,\hat{\beta}} + \alpha_{i,\hat{\alpha}} - \alpha_{i+\hat{\beta},\hat{\alpha}} = \alpha_{i,\hat{\alpha}} + \alpha_{i+\hat{\alpha},\hat{\beta}} + \alpha_{i+\hat{\alpha}+\hat{\beta},-\hat{\alpha}} + \alpha_{i+\hat{\beta},-\hat{\alpha}}.
  \label{eqn:lattice-g-tr}
\end{equation}
This transformation is locally generated by the operator, illustrated in figure \eqref{fig:g-tr},
\begin{equation}
  G_{i,\hat{\alpha}} = \sum_{\hat{\beta} \neq \hat{\alpha}} \left( E_{i,\hat{\alpha},\hat{\beta}} - E_{i,\hat{\alpha},-\hat{\beta}} \right).
  \label{eqn:gauss-law}
\end{equation}
This operator is the analog of the Gauss' law operator in lattice electrodynamics.
The full gauge-transformation operator on a lattice without a boundary, which we include for completeness, is
\begin{equation}
  \mathcal{G} [\alpha_{i,\hat{\alpha}}] = e^{i \sum_{\text{links}} \alpha_{i,\hat{\alpha}} G_{i,\hat{\alpha}}}.
  \label{eqn:g-tr-op}
\end{equation}

On a lattice with a boundary, however, we have to distinguish between large and small gauge transformations.\footnote{It is important to note here that large gauge transformations only exist when we take $B_{0i} = 0$ boundary conditions. We also need to pick one more boundary condition, but we may leave that unspecified without causing any disasters.}
Small gauge transformations are generated by operators of the form \eqref{eqn:gauss-law} on interior links, which are links for which all four adjoining faces are actually in the lattice.
Large gauge transformations are generated by operators similar to \eqref{eqn:gauss-law} on boundary links, except that we remove the $E$s that live on non-existent faces.
We will refer to generators of large gauge transformations as $G^{l}$, or just $G$ when it is obvious from context.

The basic gauge-invariant operators here are the $E$s themselves and the Wilson boxes
\begin{equation}
  W_{i} = B_{i,\hat{x},\hat{y}} + B_{i,\hat{z},\hat{x}} + B_{i,\hat{y},\hat{z}} + B_{i+\hat{x}+\hat{y}+\hat{z},-\hat{x},-\hat{y}} + B_{i+\hat{x}+\hat{y}+\hat{z},-\hat{z},-\hat{x}} + B_{i+\hat{x}+\hat{y}+\hat{z},-\hat{y},-\hat{z}},
  \label{eqn:wilson-box}
\end{equation}
which can be thought of as the sum of all $B$s around a box with the signs so that the cross product of the two directions is pointing inwards.
Note that these are invariant under both large and small gauge transformations.
The Hamiltonian, up to lattice-spacing-dependent factors, is of the form
\begin{equation}
  H = \sum_{\text{faces}} E^{2} + \sum_{i} W_{i}^{2}.
  \label{eqn:two-form-hamiltonian}
\end{equation}

\begin{figure}[h]
  \centering
  \includegraphics[width=100mm]{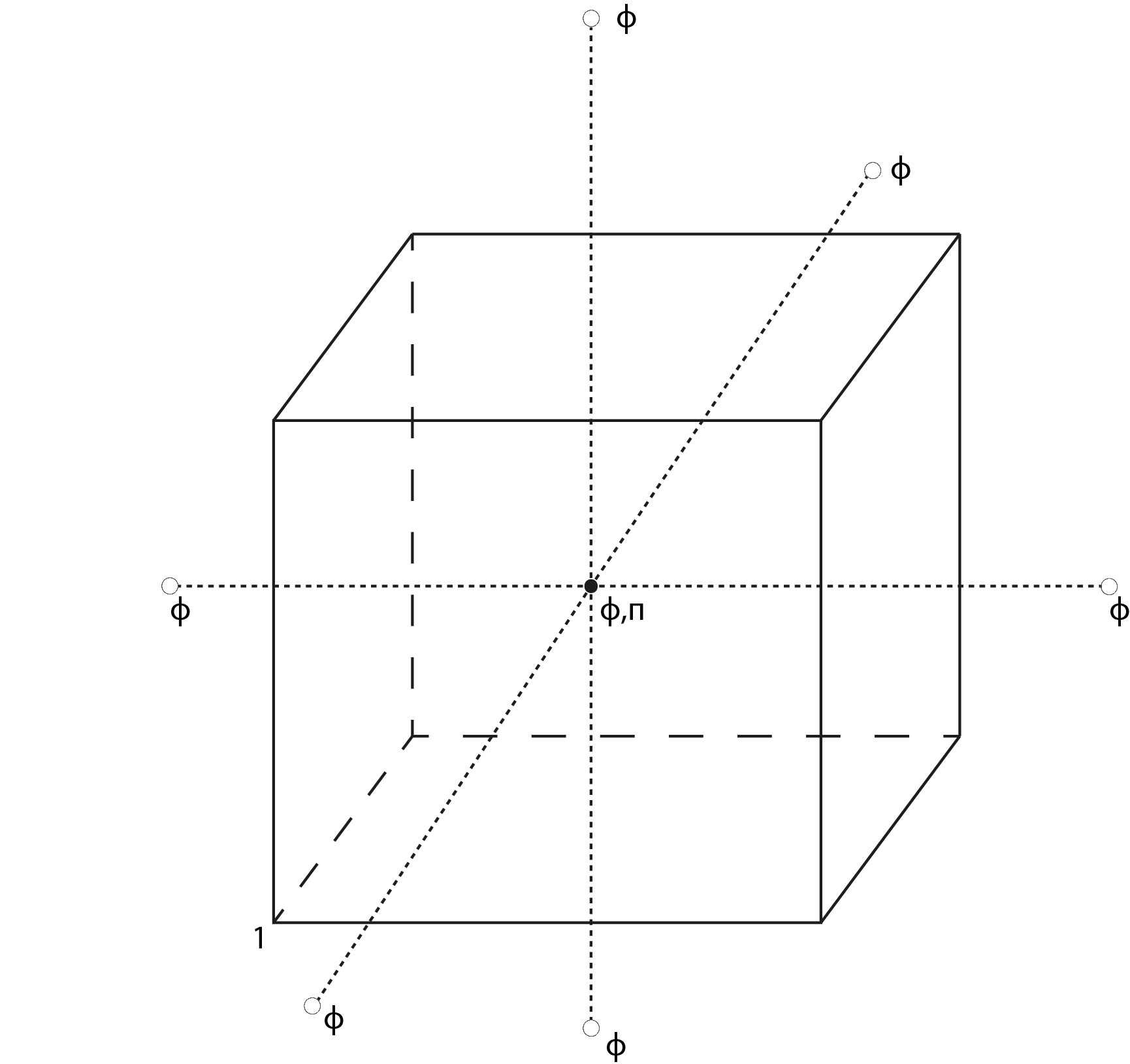}
  \caption{An illustration of Kramers-Wannier duality. The two-form fields live on the faces of the box, and the scalar field lives on the vertices of the dotted lattice. The dual vertices outside the original box, whose $\phi$ variables generate large gauge transformation, don't have a $\pi$ variable.}
  \label{fig:kw}
\end{figure}

Kramers-Wannier duality translates the two-form field to a scalar field living on the dual lattice, whose vertices are at the centers of the boxes of the original lattice and whose links pass through the faces of the original lattice, see figure \eqref{fig:kw} for an illustration.
The map between the variables is
\begin{align}
  \pi_{i} &= W_{i} \nonumber\\
  \phi_{i+\hat{\alpha}} - \phi_{i} &= E_{i+\hat{\alpha},\hat{\beta},\hat{\gamma}}, \quad (\alpha,\beta,\gamma) \text{ cyclic.}
  \label{eqn:kw-duality}
\end{align}
With this assignment, the Hamiltonian \eqref{eqn:two-form-hamiltonian} becomes
\begin{equation}
  H = \sum_{i,\hat{\alpha}} (\phi_{i+\hat{\alpha}} - \phi_{i})^{2} + \sum_{i} \pi_{i}^{2},
  \label{eqn:scalar-hamiltonian}
\end{equation}
which is the Hamiltonian of a free massless scalar field.
Note that multiplicative factors have been ignored in \eqref{eqn:kw-duality} and \eqref{eqn:scalar-hamiltonian}.

To see what the scalar duals of the large gauge transformations { generators} are, let us specialize to a lattice that is just one box, see figure \eqref{fig:kw}.
Calling one of the vertices just $1$, the large gauge transformations for such a lattice are generated by sums of two $E$s, for example
\begin{equation}
  G^{l}_{1,\hat{x}} = E_{1,\hat{x},\hat{y}} + E_{1,\hat{x},\hat{z}}.
  \label{eqn:lgt-example}
\end{equation}
The scalar dual of this is
\begin{equation}
  G^{l}_{1,\hat{x}} = \phi_{1-\hat{y}} - \phi_{1-\hat{z}}.
  \label{eqn:lgt-scalar}
\end{equation}
But, notice that these two $\phi$s sit at the centres of boxes that are not in our original lattice, and so they have no corresponding $\pi$s!
Thus, we see that large gauge transformations dualise exactly to boundary $\phi$s that have zero Poisson brackets with every other variable, which is the same property found in the previous section.

Further, let us notice that the map \eqref{eqn:kw-duality} does not uniquely fix the $\phi$ variables; if all the $\phi_{i}$s change by the same amount, the two-form variables are unaffected.
In other words, the zero-mode $\sum_{i} \phi_{i}$ is unfixed by the duality.
Taking the part of the zero-mode that has trivial Poisson brackets with every variable, we find the variable
\begin{equation}
  \sum_{i \in \partial} \phi_{i}.
  \label{eqn:boundary-zero-mode}
\end{equation}
This cannot be written purely as a sum of $G^{l}$s, as can be seen by the fact that all the $G^{l}$s are differences of boundary $\phi$s of the form \eqref{eqn:lgt-scalar} and thus insensitive to a boundary zero-mode.
Thus, we see that the zero-mode of the centre is not a gauge transformation generator, as found in the last section!

It is worth noting that this last charge is distinct from the constant shift symmetry $\delta \phi_{i} = a$ of the scalar Hamiltonian.
This symmetry is generated by the operator $\sum_{i} \pi_{i}$, as opposed to \eqref{eqn:boundary-zero-mode}, and it is only a symmetry of the Hamiltonian not an element of the centre.
In particular, the asymptotics \eqref{eqn:phi-asymptotic} rule out a constant shift in $\phi$ in the Minkowski case.
Though we used the shift symmetry to motivate the construction of the charge \eqref{eqn:boundary-zero-mode}, the facts that it is in the centre and that it cannot be written as a combination of the large gauge transformations are true independently of it.

Finally, note that if we assume the scalar theory to have the boundary centre variables as above and add interaction with another scalar field, the two-form picture drops away \emph{but the boundary variables are still in the centre}.
This could be { the reason} why we were able to reconstruct the charge from two-form large gauge transformations even in the interacting case in section \ref{sec:main-point}.

Between the fact that these boundary $\phi$s are in the centre and dual to large gauge transformations { charges}, and the fact that the zero-mode is in the centre but not dual to a large gauge transformation, we propose to interpret the asymptotic charge \eqref{eqn:soft-charge-1} as involving exactly this sort of centre variable.

\section{Conclusions}
We have shown that the charges whose conservation is the scalar soft theorems in four dimensions can be rewritten as 
generators of large gauge transformations of a Hodge-dual two-form field.
This is true for all or almost all of the charges depending on whether we consider the theory to be that of a compact or non-compact scalar.
The rewriting serves two purposes: First it gives an interpretation of the soft theorem in the general framework of the gauge symmetry/soft theorem connection, and second it allows us to find a symmetry action of these charges on the phase space consistent with the equations of motion --- the charges commute with the massless scalar everywhere.
This structure of two-form large gauge transformations being dual to scalar charges that commute with all local operators has an analog on a finite lattice, showing that it is not very exotic.

One important subtlety noted in the beginning was that in the theory of interest the soft theorem only exists at tree level, since the massless scalar generically gets a mass at loop level.
 Our attitude to this issue  was to only talk about the classical theory and therefore consider the charges as phase space variables rather than quantum operators. However, there are examples of theories --- like the $\mathcal{N} = 2$ interacting Wess-Zumino theory --- which contain vertices similar to the ones we have considered, are renormalisable, and in which the masslessness is protected by a symmetry even at loop level; in these theories, we expect the above considerations to be true for quantum operators.

Finally, we should spend a few words on interpretation: how should we think of the two-form theory?
At one level, the two-form theory is just a field redefinition of the scalar theory and therefore 
trivially correct; 
like many field redefinitions,  the resulting theory is 
non-local. This is of course not a satisfactory answer: the 
interesting question is whether we must expect other consequences of the dual gauge symmetry to be manifest in the scalar theory.
We are unfortunately unable to comment on this question at present and leave it to future work.

Even if one takes the most conservative view about the `existence' of the dual field, there 
are still  lessons to take away. The existence of 
charges and the non-triviality of their Ward identities should be reason enough to believe that the scalar theory has edge modes of the form discovered in this work. The fact that the charge \eqref{eqn:soft-charge-1} generates the non-trivial soft theorem \eqref{eqn:soft-thm} as a Ward identity means that the charge is a non-zero operator. Its symmetry action, however, is far from obvious;  the two-form theory, then, is a clever if roundabout way of understanding the symmetry action of these charges.
The existence of the auxiliary two-form theory should also give us confidence that the commutation relations we find are consistent, since they follow from obvious transformation rules in the two-form theory. This perspective is similar to recent 
understanding \cite{Hosseinzadeh:2018dkh,Freidel:2018fsk} of  the two polarisations of  soft photons in QED in terms of electric and magnetic large gauge transformations.

\acknowledgments
We thank William Donnelly, Madhusudhan Raman and especially Alok Laddha for various discussions, and Shivani Upadhyaya for making figure 2; RMS would also like to thank Upamanyu Moitra and Sandip P. Trivedi for collaboration on related work \cite{Moitra:2018lxn}. MC would like to thank Perimeter Institute and specially Bianca Dittrich for hosting him under a visiting researcher program in Fall 2017.  MC is partially supported by PEDECIBA and SNI of Uruguay. FH is supported by a Vanier Canada Graduate Scholarship. RMS would like to thank the Perimeter Institute for hosting him under their ``Visiting Graduate Fellows'' program; the DAE, Government of India for supporting research at TIFR; and the Infosys Endowment for Research into the Quantum Structure of Spacetime. Many fruitful discussions with Alok Laddha were carried out in the International Centre for Theoretical Sciences (ICTS) program ``Kavli Asian Winter School (KAWS) on Strings, Particles and Cosmology 2018'' (Code: ICTS/Prog-KAWS2018/01). Research at Perimeter Institute is supported by the Government of Canada through the Department of Innovation, Science and Economic Development Canada and by the Province of Ontario through the Ministry of Research, Innovation and Science.

\appendix
\section{Conventions} \label{sec:conventions}
In this appendix, we write down the factors in our conventions for exterior derivatives, Hodge stars, etc.

The components of a $p$-form $\omega$ are given by
\begin{equation}
  \omega = \frac{1}{p!} \omega_{\mu_{1}\ldots \mu_{p}} dx^{\mu_{1}} \cdots dx^{\mu_{p}}.
  \label{eqn:p-form-comps}
\end{equation}
Deifning the anti-symmetrisation brackets with a normalisation, so that for the Levi-Civita tensor
\begin{equation}
  \varepsilon_{[\mu_{1}\ldots\mu_{p}]} = \varepsilon_{\mu_{1}\ldots\mu_{p}},
  \label{eqn:anti-symm-bracket-def}
\end{equation}
The exterior derivative is given by
\begin{align}
  d \omega &= \frac{1}{p!} \partial_{\nu} \omega_{\mu_{1}\ldots\mu_{p}} dx^{\nu} dx^{\mu_{1}} \cdots dx^{\mu_{p}} \nonumber\\
  &= \frac{1}{p!} \partial_{[\nu} \omega_{\mu_{1}\ldots\mu_{p}]} dx^{\nu} dx^{\mu_{1}} \cdots dx^{\mu_{p}}.
  \label{eqn:d-defn}
\end{align}
The components of the exterior derivative are
\begin{equation}
  (d \omega)_{\mu_{1}\ldots\mu_{p+1}} = (p+1) \partial_{[\mu_{1}} \omega_{\mu_{2}\ldots\mu_{p+1}]},
  \label{eqn:d-comps}
\end{equation}
which is the natural factor when $\omega_{\mu_{2}\dots\mu_{p+1}}$ is completely antisymmetric.

The Hodge star of a $p$-form is
\begin{equation}
  * \omega = \frac{1}{p! (4-p)!} \varepsilon_{\mu_{1} \ldots \mu_{4}} g^{\mu_{1}\nu_{1}}\cdots g^{\mu_{p}\nu_{p}} \omega_{\nu_{1}\ldots\nu_{p}} dx^{p+1}\cdots dx^{4},
  \label{eqn:hodge-defn}
\end{equation}
which in component language is
\begin{equation}
  (* \omega)_{\mu_{p+1}\ldots\mu_{4}} = \frac{1}{p!} \varepsilon_{\mu_{1} \ldots \mu_{4}} g^{\mu_{1}\nu_{1}}\cdots g^{\mu_{p}\nu_{p}} \omega_{\nu_{1}\ldots\nu_{p}}.
  \label{eqn:hodge-comp}
\end{equation}
Here,
\begin{equation}
  \varepsilon_{0123} = \sqrt{|g|},
  \label{eqn:levi-civita-tensor}
\end{equation}
so that it is a tensor and the $4$-form associated to it by \eqref{eqn:p-form-comps} is the volume form.
With this convention,
\begin{equation}
  \omega \wedge *\omega = \frac{1}{p!} \omega_{\mu_{1}\ldots\mu_{p}} \omega^{\mu_{1}\ldots\mu_{p}} \sqrt{|g|} d^{4} x.
  \label{eqn:forms-ip}
\end{equation}

\bibliographystyle{JHEP}
\bibliography{refs}

\end{document}